\documentclass[jeosman,preprint,fleqn,showpacs,showkeys]{revtex4}
\usepackage{graphicx}
\usepackage{amssymb,amsmath,bm}
\pdfoutput=1

\begin{document}

\title{Double pendulum model for tennis stroke including a collision process.}

\author{Sun-Hyun Youn\footnote{email: sunyoun@chonnam.ac.kr, fax: +82-62-530-3369}}
\address{Department of Physics, Chonnam National University, Gwangju 500-757, Korea}

\begin{abstract}

 By means of adding a collision process between the ball and racket in double pendulum model,
 we analyzed the tennis stroke.
 It is possible that the speed of the rebound ball does not
 simply depend on the angular velocity of the racket, and higher
 angular velocity sometimes gives lower ball speed. We numerically showed that
 the proper time lagged racket rotation increases the speed of the rebound ball by $20 \%$.
 We also showed that the elbow should move in order to add the angular velocity of the racket.



\end{abstract}


\maketitle

\section{Introduction}

 The double pendulum has been studied  as an example of
   the chaotic motion in physics\cite{chaos1,chaos2,chaos3}. If we
   consider the first cycle of the pendulum, the double pendulum model
   holds application in sports such as golf\cite{Ref3,Ref4,Ref5,Ref6},
   baseball\cite{RodCross05}, and tennis\cite{RodCross11}. Since
   the double pendulum is not a simple linear system, the motion of
   the pendulum can not be optimized as a simple analytic form.
    The  swing pattern utilised to maximize the angular velocity of the hitting rod
such as racket, bat, and club has been analyzed on the assumption
that the angular velocity is the dominant factor for the speed of
the rebound ball\cite{RodCross11}.

    For a tennis stroke, the two rods for double pendulum are
    an arm and a forearm  for first rod and  a racket for the second
    rod. In our model, we added the collision process between the
    ball and the racket. Without the collision process\cite{RodCross11}, there is no
    criteria  to attain  high speed of the rebound ball except
    for the angular speed of the second rod, racket. If we set the impact angle of the first
     rod at which the ball hits the racket,  the speed of the rebound ball is mainly dependent on the angular
    velocity of the second rod.   On the other hand, if we release the impact
angle of the first rod,  the speed of the rebound ball does not
remain as a simple function of the angular speed of the hitting rod.
We showed that the speed of the rebound ball is different even
though the angular velocities of the second rod at the contact time
are same. Furthermore, considering the whole stroke, the maximum
angular velocity for the lower speed of the rebound ball is greater
than the maximum angular velocity  for the higher speed of the
rebound ball. Therefore, to get  maximum speed of the rebound ball,
it's not sufficient to set the condition to generate high angular
velocity of the hitting racket.

 The collision between the racket and ball has been studied in
 various ways \cite{RodCrossImpact,ObliqueImpact}. In our simple collision model,
 we assumed that the racket is a simple one dimensional rod without any
 nodal motion and the collision occurs in one dimension. This
 assumption is valid  if the racket and a ball moves in the same
 line for the short collision time. Although, our model does not give any
 detailed information on the collision about the effect of the tension of the string and the mass
 distribution of the racket, however provides some insights on
 the proper swinging pattern  to get  maximum speed of the rebound ball

 In this article, we also analyzed the tame lagged torque
  effect for the double pendulum system. By applying time independent constant torques on first rod and the second rod,
  the speed of the rebound ball can be calculated for  certain initial conditions.
  In the same condition, if we simply  hold the racket for a short time without enforcing a torque
  and with subsequent application of torque, the naive intuition
  estimates decrease in the speed of the rebound ball.  The reason behind this is
  application of less energy for the double pendulum system. However,
  the speed of the rebound ball increase by $20 \%$ by choosing
  the proper delay time.  It's mainly because the double pendulum
  system is not a simple linear system. Adding energy to the
  double pendulum system does not directly increase the speed of the
  rebound ball. We also analyzed movement of the elbow  to which the first rod of the
double pendulum is  attached. At a first glance, if we add extra
movement towards the rebound ball's direction at contact time, the
speed of the ball  increases; however, with varying results.
Apparently, it becomes clear that, the double pendulum system is
really a nonlinear system.

  The present paper is organized as follows: In Section II, we
 introduce a double pendulum system including the collision process.
 In Section III, the differential equations obtained in section II are solved, we numerically
 showed that the speed of the rebound ball does not simply depend
 on the angular speed of the racket. For some cases, the higher
 angular velocity gives lower speed of the rebound ball.
 In Section IV, we analyzed the dependence of the racket mass and
 length of the first rod. The general properties of swing system
 has been demonstrated.  When we applied  a time dependent torque on the racket,
 we could increase the speed of the rebound ball. The time lagged
 rotation of the racket was analyzed in section V. In section VI,
 the elbow movement is analyzed to add additional speed to the ball.
 In Section VII, we
 summarize the  main results and discuss the application of our results.

\section{Racket system including the collision with a ball}

\begin{figure}[htbp]
\centering
\includegraphics[width=5cm]{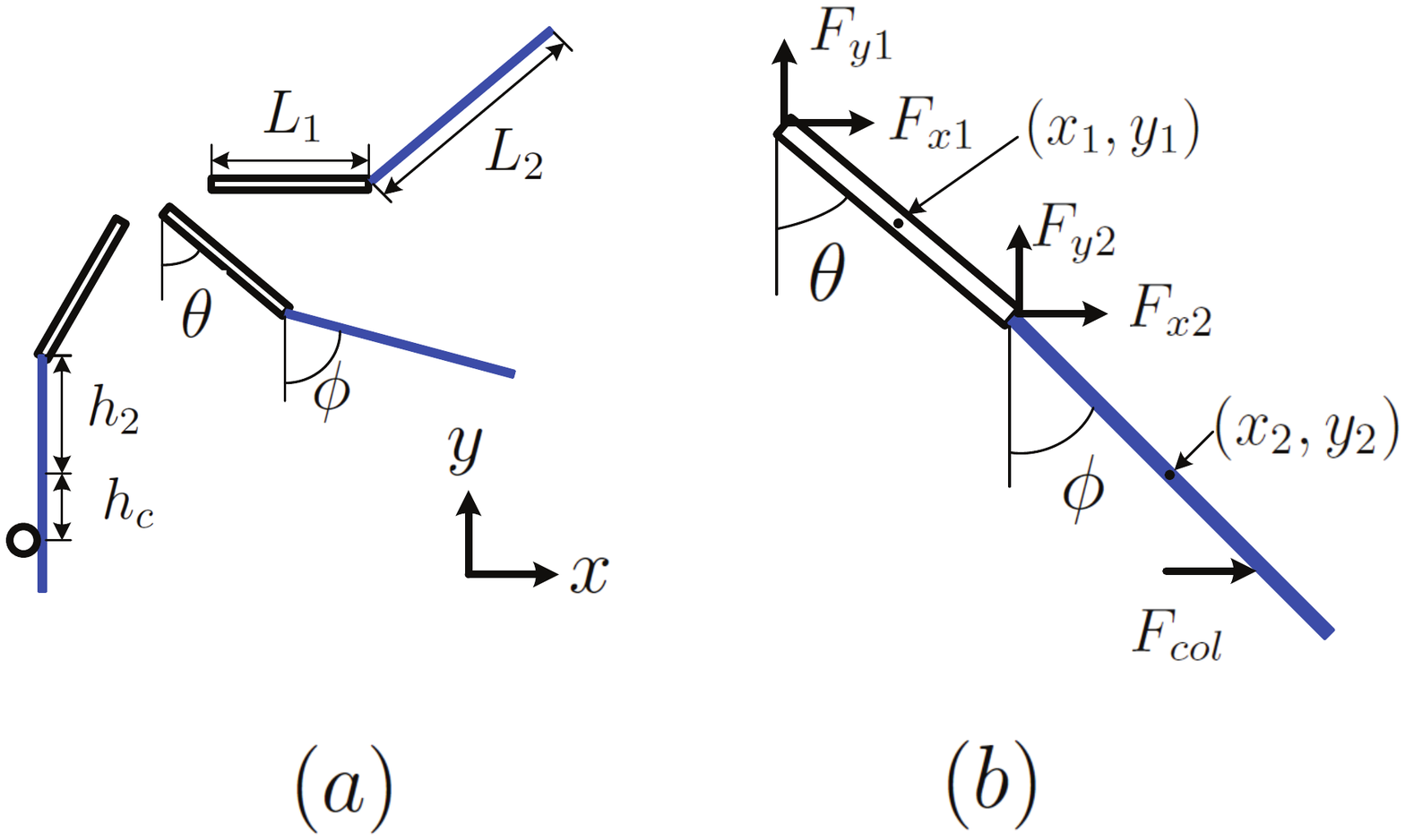}
\caption{Schematic diagram of arm and racket system. (a) Length of
the first rod (forearm and arm system) is $L_1$, and the length of
the racket is $L_2$. The angles $\theta$ and $\phi$ are defined from
the $-y$ direction. (b) The Force on two rods. $(F_{x1},F_{y1})$ is
the force acted on the first rod at the joint between elbow and the
first rod.  $(F_{x2},F_{y2})$ is the force acted on the second rod
at the joint between the two rods. $F_{col}$ is the  force on the
second rod by a ball. } \label{FigRacket}
\end{figure}

The geometry of the double pendulum model for the swing of a racket
is shown in Fig \ref{FigRacket}-(a). Though, this geometry is for
the left-handed player if we see from the $+z$ direction, it's
originally  related to the real double pendulum problem in the
gravitational field \cite{RodCross05}. Our basic model and some
notations are closely related with those in work reported by  Rod
Cross \cite{RodCross11}.  The elbow moves in the $xy$ plane and the
arm and the racket also moves in $xy$ plane.  We also modeled the
arm and the forearm as a  simple uniform rod with mass $M_1$ and
length $L_1$. The racket including the hands is also treated as a
uniform rod with mass $M_2$ and length $L_2$. The arm and the racket
rotate in a clockwise direction in a plane at  angular velocities
$\omega_1 = -d \theta /d t $ and $\omega_2 = - d \phi / dt$,
respectively. If we assume that the velocity of the  elbow is $
(V^E_{x} , V^E_{y} )$ then the velocities of the center of the first
rod (arm and forearm system) and the second rod (hand and racket
system) becomes
\begin{eqnarray}
v_{x1}&=& V^E_{x}  - h_1 \omega_1  \cos  \theta
\nonumber \\
v_{y1}  &=& V^E_{y}  - h_1 \omega_1  \sin  \theta
\nonumber \\
v_{x2} &=& V^E_{x}- L_1 \omega_1 \cos  \theta  - h_2 \omega_2 \cos
\phi
\nonumber \\
v_{y2}  &=& V^E_{y} - L_1 \omega_1  \sin  \theta  - h_2 \omega_2
\sin \phi  \label{EqA1}
\end{eqnarray}
,where $(v_{x1},v_{y1})$ , $(v_{x2},v_{y2})$  are the velocity of
the center of the mass of the first rod and the second rod,
respectively. The center of
 masses are located in the middle of the rod,   $h_1=L_1 /2 $ and
 $h_2 = L_2 / 2$, since we assumed uniform rods.

Let the force from the elbow  to the first rod on joint point
between the elbow and the first rod be $(F_{x1}, F_{y1})$, and the
force from the first rod on joint point between two rods  be
$(F_{x2}, F_{y2})$, then the equations of the motion for two center
of  masses become
\begin{eqnarray}
M_1 \frac {dv_{x1}}{dt} =  F_{x1} - F_{x2} \nonumber \\
M_1 \frac {dv_{y1}}{dt} =  F_{y1} - F_{y2}   \label{EqF1}
\end{eqnarray}
\begin{eqnarray}
M_2 \frac {dv_{x2}}{dt} &=&   F_{x2} + F_{col}\nonumber \\
M_2 \frac {dv_{y2}}{dt} &=&   F_{y2}   \label{EqF2}
\end{eqnarray}
The  two forces $(-F_{x2}, -F_{y2})$ are reaction force of $(F_{x2},
F_{y2})$ from the second rod. We added the force $F_{H}$ from the
ball to the second rod at the point of $h_H$ from the center of the
mass.  This force from the collision between the racket and the ball
changes  the torque equation in \cite{RodCross11}  and the
two torque equations becomes %
\begin{eqnarray}
I_{2,cm} \frac { d \omega_2}{dt} = C_2 + F_{x2} h_2 \cos \phi +
F_{y2} h_2 \sin \phi - F_{col} h_{H} \cos \phi   \label{Torque2}
\end{eqnarray}
\begin{eqnarray}
I_{1,cm} \frac { d \omega_1 }{dt} &=& C_1 - C_2 + F_{x1} h_1
\cos \theta \nonumber \\
 &+& F_{y1} h_1 \sin \theta + F_{x2} (L_1 - h_1) \cos
\theta + F_{y2}(L_1 -h_1 ) \sin \theta \label{Torque1}
\end{eqnarray}
 ,where $C_1$ is the torque on joint point between the elbow and the first rod and $C_2$ is the
 torque on the joint point between the two rods.
From the Eqs. \ref{EqF1} - \ref{Torque1}, we obtain two equations
for the time derivative of two angular velocities as follows
\begin{eqnarray}
 \frac { d \omega_1 }{dt} &=&  \frac{P (I_{2,cm}+ M_2 h_2^2 ) - Q M_2 h_2 L_1 \cos(\phi-\theta)+S_1 }
 {(I_{2,cm} + M_2 h_2^2 )(I_{1,cm} + M_1 h_1^2 + M_2 L_1 ^2)  - M_2^2 h_2^2 L_1^2 \cos^2 (\phi-\theta)  } \nonumber \\
  \frac { d \omega_2 }{dt} &=&  \frac{Q (I_{1,cm} + M_1 h_1^2 + M_2 L_1 ^2 ) - P M_2 h_2 L_1  \cos(\phi-\theta)+S_2}
  {(I_{2,cm} + M_2 h_2^2 )(I_{1,cm} + M_1 h_1^2 + M_2 L_1 ^2) - M_2^2 h_2^2 L_1^2 \cos^2 (\phi-\theta) }
 \label{Omega12}
\end{eqnarray}
where $a_x  = \frac{d V_x^E}{dt} $, $a_y  = \frac{d V_y^E}{dt} $,
and
\begin{eqnarray}
 P &=& C_1 - C_2 -  M_2 h_2 L_1 \omega_2 ^2 \sin (\phi-\theta) + ( M_1
 h_1 + M_2 L_1)  [ a_x \cos \theta + a_y \sin \theta ] \nonumber \\
  Q &=& C_2 + M_2 h_2 L_1  \omega_1 ^2 \sin (\phi-\theta) + M_2 h_2
 [ a_x \cos \phi + a_y \sin \phi ]
 \label{EqPQ}
\end{eqnarray}
$P$ and $Q$ are similar to the results in \cite{RodCross11} with
setting $g=0$, and the collision force from the ball add following
two terms
\begin{eqnarray}
 S_1  &=& F_{col} L_1 [( 2 I_{2,cm} + h_2 (h_2 -h_H ) M_2 ) \cos \theta - h_2 (h_2  + h_ H ) M_2 \cos (\theta-2 \phi)]
 \nonumber \\
S_2  &=& F_{col} [ (2 h_H ( I_{1,cm} + M_1 h_1 ^2 + M_2 L_1 ^2) \nonumber \\
&+& h_2 ( 2 I_{1,cm} + 2 M_1 h_1 ^2 + M_2 L_1 ^2 )) \cos \phi - h_2
M_2 L_1 ^2 \cos (2 \theta - \phi) ]  \label{EqS12}
\end{eqnarray}
The collision force defines the motion of the ball as follow ,
\begin{eqnarray}
 m_b \frac{ d^2  x_b }{dt^2}   &=&  F_{col} \nonumber \\
  &=& - f_k (x_H - x_b) \\ \label {Eqfk}
x_H &=& x_E + L_1 \sin \theta + (h_2 + h_H ) \sin \phi \label {EqxH}
\end{eqnarray}
,where  $x_H$ is the $x$ component of the hitting part of the racket
which locates $h_2 + h_H$ from the bottom of the racket, and $x_E$
is the $x$ component of the joint between the elbow and arm.

 The force
between the racket and the ball should be repulsive, so that  the
force form becomes
\begin{eqnarray}
  f_k (x_H - x_b) &=& k (x_H - x_b ) ~~~~~     if~~ x_H > x_b  \nonumber \\
  &=& 0~~~~~~~~~~  if~~ x_b \geq x_H  .  \label {Eqfkform}
\end{eqnarray}
If we consider this harmonic force between the ball and the second
rod, the period of the oscillation is \cite{RodCross00}
\begin{eqnarray}
  T = \frac {2 \pi}{ \sqrt{\frac{k (m_b + M_2 )}{m_b M_2}}}.   \label {Period}
\end{eqnarray}
  Since the half of this period is the collision duration between the ball
and the racket, we controlled  the $k$  value  from $188 N/m$ to
$19000 N/m$. Subsequently, the collision time varies from $50ms$ to
$5ms$. However, if the contraction length is large, the Hook's model
is not valid for the ball and the racket system.  Then the force can
be rewritten as $ k x_m  \sin ( \frac{\pi x} {x_m}) $
\cite{RodCross00}. or  $ \frac{2 k x_m }{\pi} \tan ( \frac{x \pi}{2
x_m} )$ in order to limit the maximum contraction length. However,
in our numerical calculation we restricted our system in order to
follow the Hook's rule.

  In this article, we assumed a simple model for the ball and racket
collision.  In our model, the ball is assumed to hit the racket when
the racket   is parallel to $y$ axis ($\phi = 0$). In addition to
this, the ball is assumed to be moving in   $x$ axis. In this case,
we numerically calculated $\theta(t) $ and $\phi(t)$ till the time
$t_0$, when  the racket is parallel to the $y$   axis
$(\phi(t_0)=0)$. With these numerical results, we set new   initial
conditions just before the collision. We assumed that the collision
occurs at $t= t_0 - T/2$, where $T$ is the period of   the  harmonic
oscillator system between the racket and the ball.   At this time,
we numerically solved the ball and two rod system attached to the
moving elbow numerically till the ball is rebound from the racket.
We determined the speed of the ball at that moment.
  Based on our simple model, the coupling constant $k$ determines
the collision duration, but the speed of the rebound ball is not
altered very much. The reason for such a behavior is consideration
of one dimensional collision. However, the purpose of this article
was to find the optimum path for the swing of the racket, we did not
extend our model to  two dimension. We only restricted our numerical
conditions to render presence of our setups in the valid region.

\section{Different ball speed with the same angular velocity of the racket}

We assumed that an arm and a forearm forms  a simple rod and the
moment of inertia about the center of mass is $I_{1,cm} = M_1 L_1 ^2
/ 2 $. The moment of inertia about the center of mass of the
hand-racket system is also assumed as $I_{2,cm} = M_2 L_2 ^2 / 2 $.
Although this assumptions are not enough to study the tennis stroke
in detail, we focused our attention on the double pendulum model,
which has its application in the tennis stroke. In this section, the
torques applied to the first forearm system $C_1$ and to the racket
system $ C_2 $ were set by $25 {N}$ and $2.5 {N}$ as described in
Rod Cross work \cite{RodCross11}. The  velocity ($m/s$) of the elbow
is assumed as follows \cite{RodCross11} ,
\begin{eqnarray}
  v^E_x (t) = -33 t + 69 t^2   \nonumber \\
  v^E_y (t) = -42 t + 174 t^2 . \label {EqVE}
\end{eqnarray}

\begin{figure}[htbp]
\centering
\includegraphics[width=5cm]{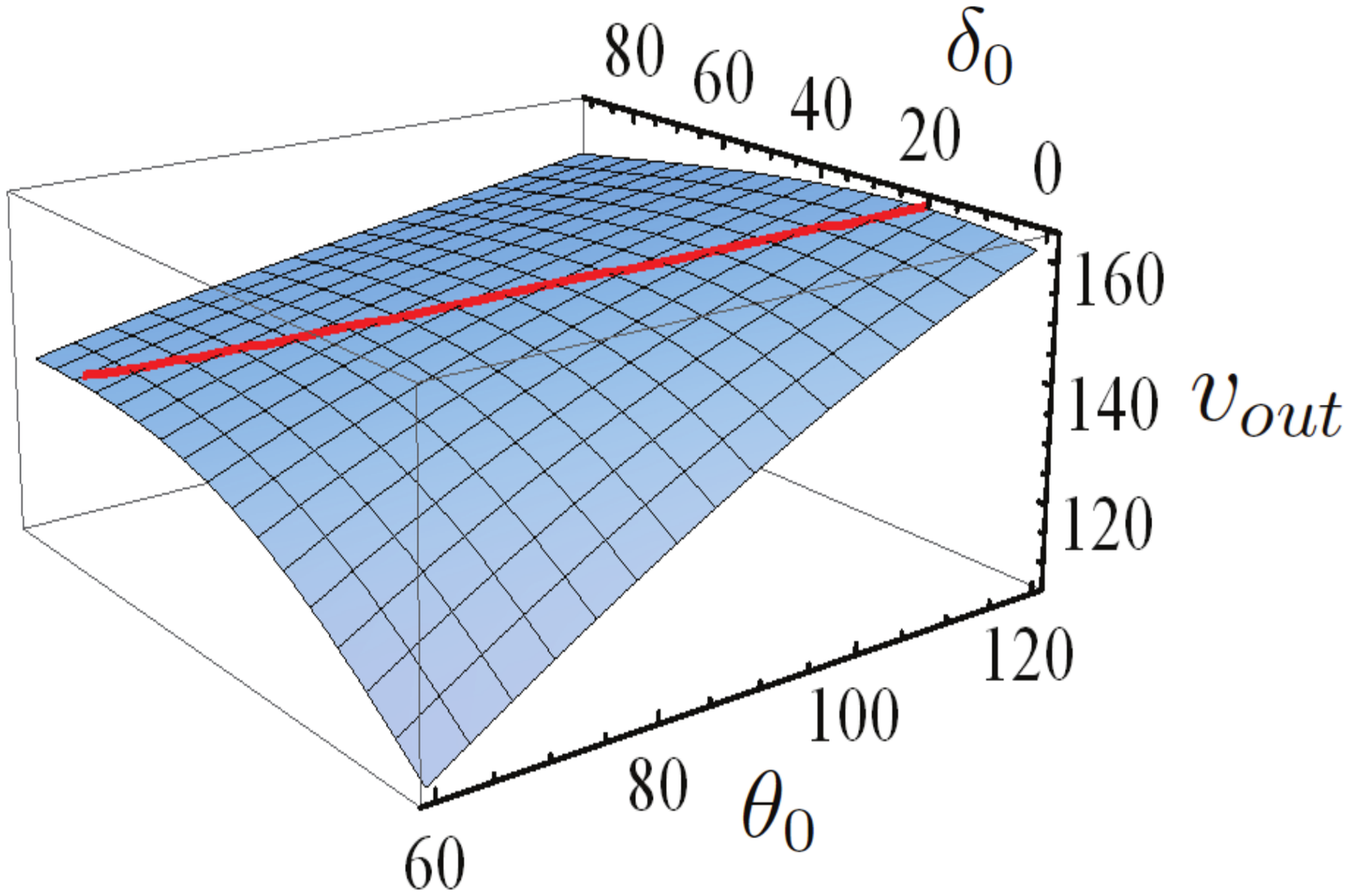}
\caption{The speed of the ball in $km/h$ unit  as a function of the
initial angle of the racket and arm with respect to the $-y$ axis.
$\theta_0$ : initial angle of the arm. $\delta_0$ : initial angle
between the racket and the first rod. The length of the first rod is
$0.3m$. Red line indicates two initial angles which gives the
maximum speed. } \label{FThetaDeltaLp3}
\end{figure}

\begin{figure}[htbp]
\centering
\includegraphics[width=5cm]{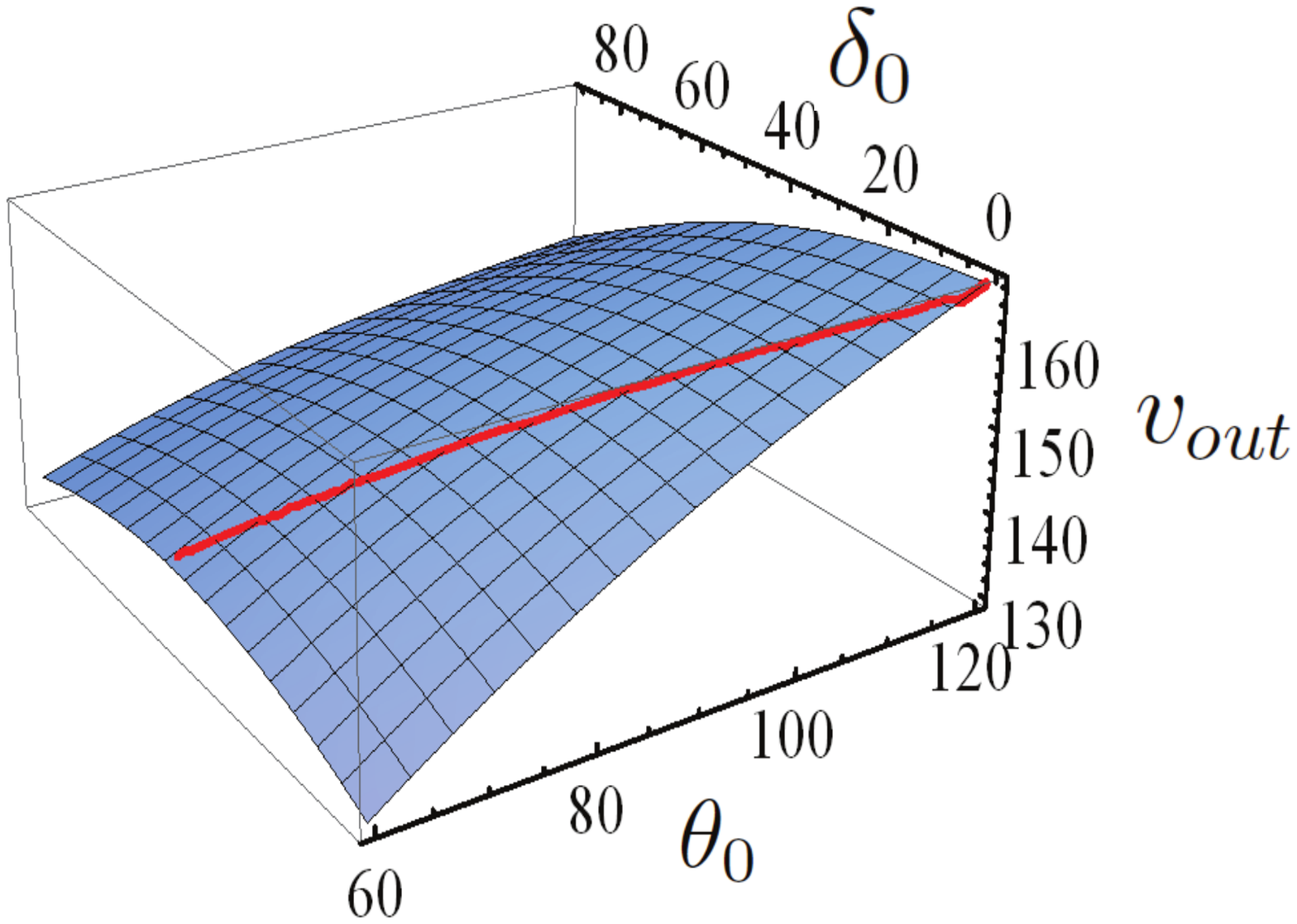}
\caption{The speed of the ball in $km/h$ unit  as a function of the
initial angle of the racket and arm with respect to the $-y$ axis.
$\theta_0$ : initial angle of the arm. $\delta_0$ : initial angle
between the racket and the first rod. The length of the first rod is
$0.4m$. Red line indicates two initial angles which gives the
maximum speed } \label{FThetaDeltaLp4}
\end{figure}

\begin{figure}[htbp]
\centering
\includegraphics[width=5cm]{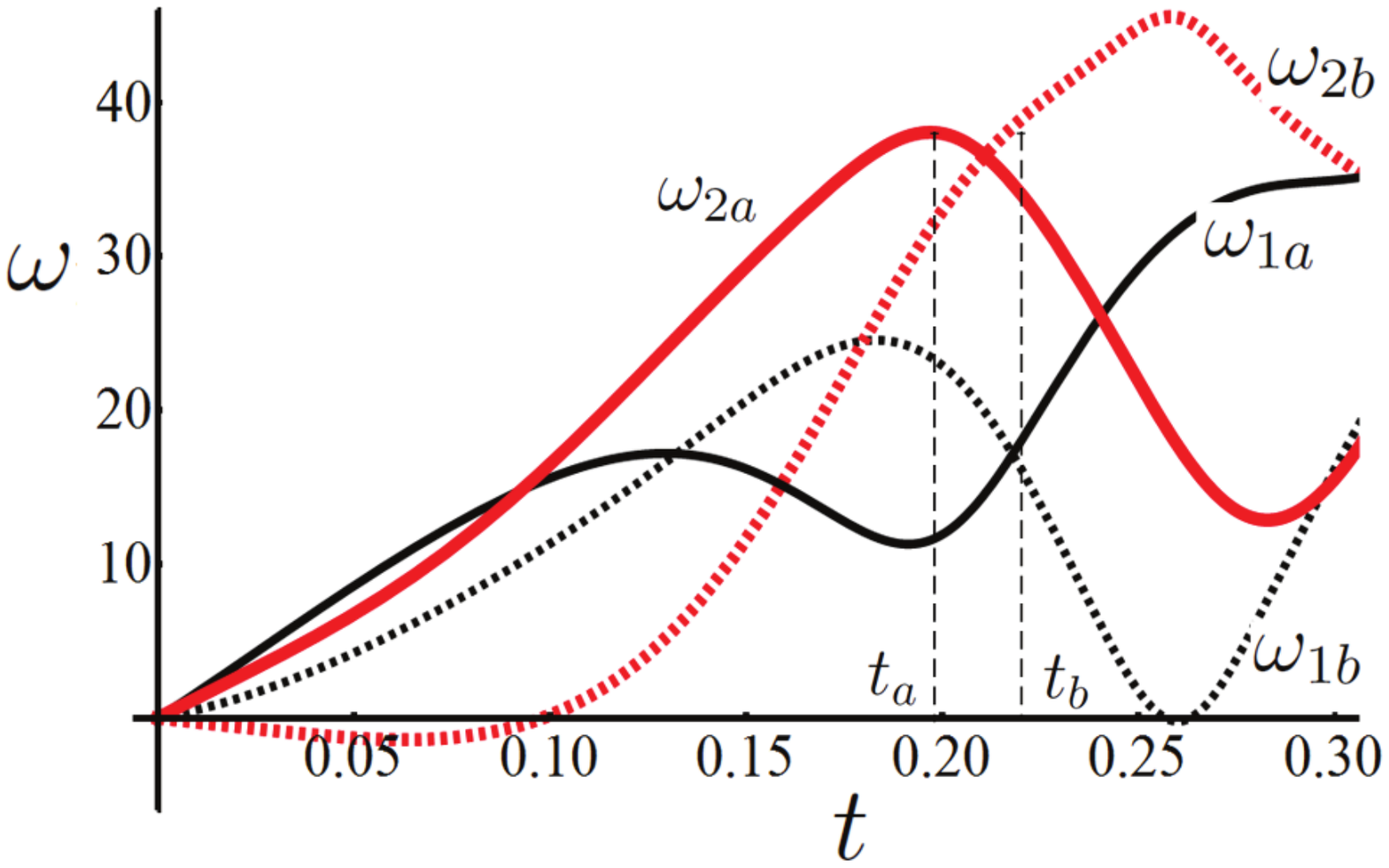}
\caption{The angular velocities $\omega_1 $ and $\omega_2$ for two
cases. The initial angle $(\theta_0, \delta_0)$ is  $(120.0
^{\circ}, 81.5 ^{\circ})$ for the case a) and $(75.0 ^{\circ},39.0
^{\circ})$ for the case b). $t_a$ and $t_b$ are the contact time.}
\label{Fw1Nw2}
\end{figure}

In Fig. \ref{FThetaDeltaLp3}, we plotted the speed of the rebound
ball from the racket as a function of the initial angles $(\theta_0
, \delta_0 )$. The angle $\delta \equiv \phi- \theta$ is defined as
the angle between the first rod and the racket. $\delta_0$ is the
initial angle at $t=0$. We assumed that the length of the first rod
(an arm and a forearm system) as $L_1 = 0.3 m $ and the mass of the
first rod as $M_1 = 2.0 kg $. And the length of the racket system as
$L_2 = 0.7$ and the mass of the racket system as $M_2=0.3kg$.
 We assumed that the initial velocity of the ball just before the
contact is $5 m/s $ to the positive $x$ direction. The $x_H$ , a
hitting position from the center of mass assumed to be as $0.15m$.
The length of the first rod (an arm and forearm system) assumed to
be as $0.3m$. This length is projection length in $x-y$ planes. We
plotted initial conditions which gives maximum speed of the rebound
ball. As $\theta_0$ changes from $60 ^{\circ}$ to $120 ^{\circ}$,
the $\delta_0$ becomes smaller and reaches $20 ^{\circ} $.
 In Fig. \ref{FThetaDeltaLp4}, we also  plotted the speed of the rebound ball
from the racket as a function of the initial angles $(\theta_0 ,
\delta_0 )$.  We only changed the length of the first rod, in other
words we extended the distance between the elbow and the hand. At
this time the $\delta_0$ is smaller as compared to the case when
$L_1 = 0.3 m $. When the initial angle $\theta_0$ is $120 ^{\circ}$,
the racket and the first rod should be in the same line so as to get
the maximum speed of the ball.
  In our model, the speed of the rebound ball is not a simple
function of the angular velocity $\omega_2$.  In Fig. \ref{Fw1Nw2},
we plotted the angular velocities $\omega_1 $ and $\omega_2$ for two
cases. For the case a) the initial angle $(\theta_0, \delta_0)$ is
$(120.0 ^{\circ}, 81.5 ^{\circ})$ and the contact time $t_a = 0.198
s$ and the angular velocity is $\omega_1 = 11.34 rad/s$.  For the
case b), the initial angle $(\theta_0, \delta_0)$ is $(75.0
^{\circ}, 39.0 ^{\circ})$ and the contact time $t_b = 0.220 s$ and
the angular velocity $\omega_1 = 17.9 rad/s$. However, the angular
velocities $\omega_2$ at the contact times have the similar value as
$37.98 rad/s$. Furthermore, the angular velocity $\omega_{2b}$
increases even after the collision time.  The interesting thing is
that the speed of the rebound ball for the case b) is smaller than
the speed of the ball for the case a). The speeds are $156km/h$ and
$141km/h$, respectively.

If we only check the angular velocity $\omega_2$, we may conclude
that the case b) gives higher speed of the rebound ball. But the
angular velocity is not the entire factor to determine the speed of
the ball.  This can be explained if we examine the angle
$(\theta_c)$ of the first rod when
 the racket contacts the ball. $\theta_c$'s are $11.5 ^{\circ}$ and
 $87.1  ^{\circ}$ for the cases a) and b), respectively. The speed of the ball
 ($v_{out}$) is also a functions of $\theta_c$ as well as $\omega_1$ and $\omega_2$.
 In order to get the maximum speed of the rebound ball, the double
 pendulum system should be examined  as a whole system.

\section{Dependence of the racket mass and the distance between the elbow and the hand}

\begin{figure}[htbp]
\centering
\includegraphics[width=5cm]{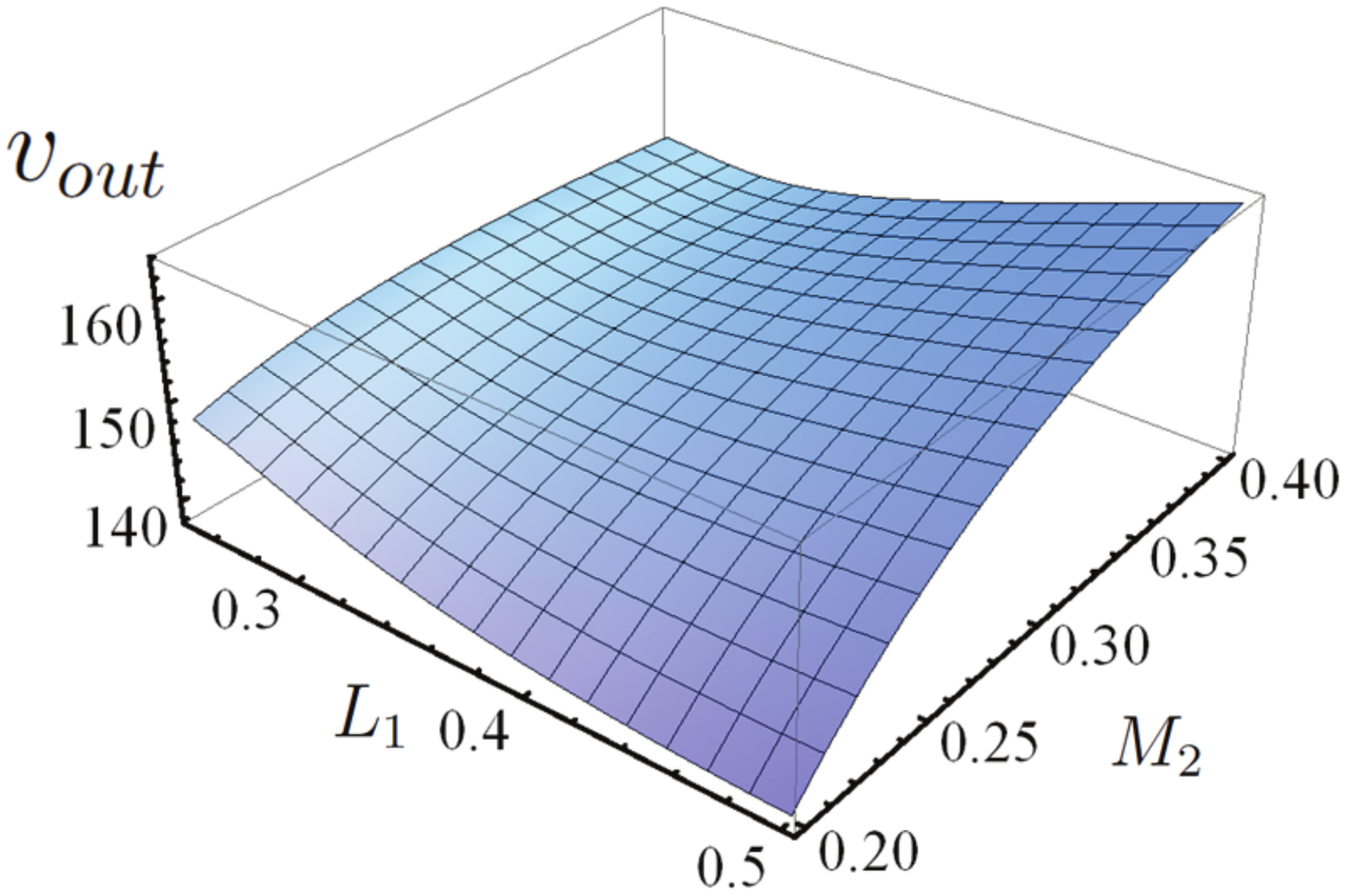}
\caption{The speed of the rebound ball  from the racket as a
function of the racket mass $(M_2)$ and the projected length in $xy$
plane  between the elbow and the hand $L_1$. The unit of $v_{out}$
is $km/h$ } \label{FLxMyVout}
\end{figure}

The mass of the racket-hand system is assumed to be  $0.3kg$, but
the mass of the racket may be changed. All the movements are assumed
to be in the $xy$ plane in our model, the projected length between
the elbow and the hand in $xy$ plane can be changed by controlling
the angle between the forearm and arm in actual tennis stroke. In
Fig. \ref{FLxMyVout}, we plotted the speed of the rebound ball from
the racket as a function of the racket mass $(M_2)$ and the
projected length in $xy$ plane  between the elbow and the hand
$L_1$. If the mass of the racket is about $0.2 {kg}$, the speed of
the rebound ball is decreased as the projected length between the
elbow and the hand is increased. However, if the racket mass is
getting heavier, the speed of the rebound ball increases. Since an
actual mass of the tennis racket is around $0.3 {kg}$, and $L_1$ is
limited, the speed of the ball is restricted.

\begin{figure}[htbp]
\centering
\includegraphics[width=5cm]{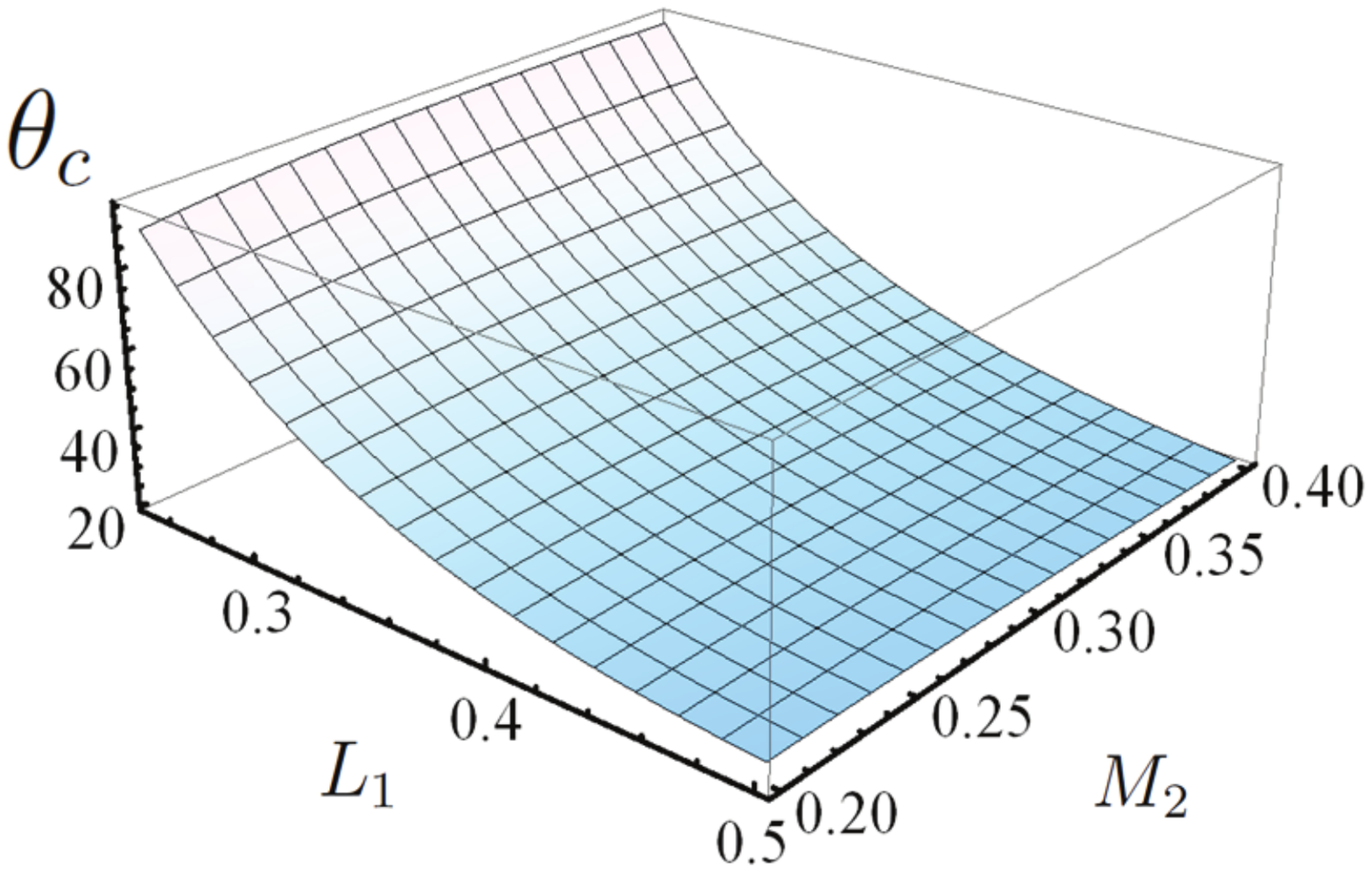}
\caption{The angle of the forearm system at the impact time as a
function of the racket mass $(M_2)$ and the projected length in
$x-y$ plane  between the elbow and the hand $L_1$.}
\label{FLxMyTheta}
\end{figure}

In Fig. \ref{FLxMyTheta}, we plotted the angle ($\theta_c$) of the
forearm system at the impact time as a function of the racket mass
$(M_2)$ and the projected length in $xy$ plane  between the elbow
and the hand $L_1 $ Regardless of the racket mass ($M_2$), the
forearm angle $\theta_c$ decrease to $20 ^{\circ}$ as the length
$L_1 $ increase.  This results shows that for the player with folded
arm whose effective projected length of the forearm system is small,
the contact angle ($ \theta_0$) should be about $90 ^{\circ}$. If
the length $L_1$ and  the contact angle $ \theta_0 $ are reduced the
ball should be hit relatively close to the body. This tendency may
explain the impact point of the ball in tennis stroke and golf.


\section{Time Lagged Torque vs. Constant Torque.}

\begin{figure}[htbp]
\centering
\includegraphics[width=5cm]{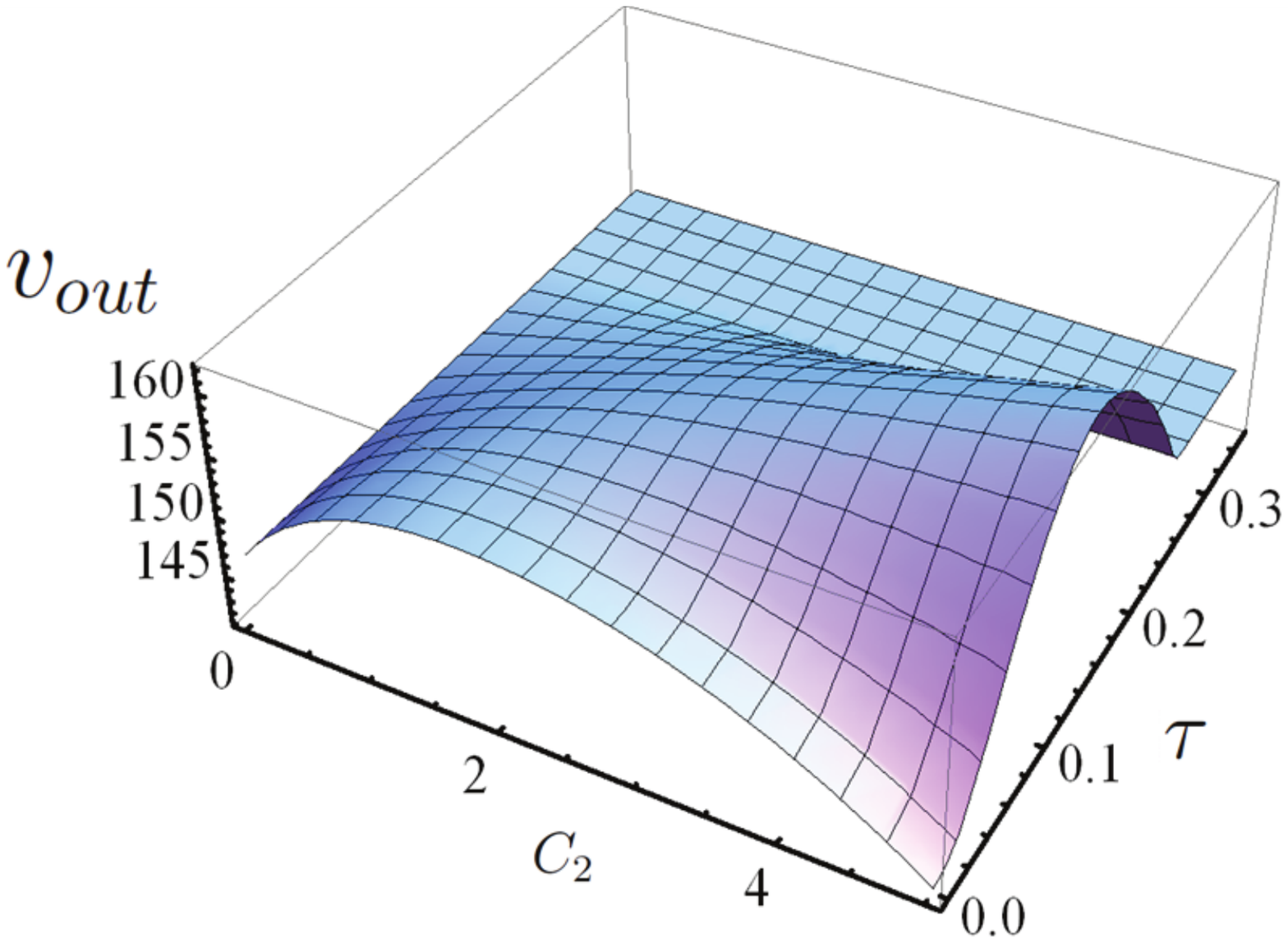}
\caption{The speed of the rebound ball as a function of the torque
 on the racket $(C_2 )$ and the time lag $\tau$. The initial
 angle  $\delta_0 $ is   $ 26.1 ^{\circ} $, the
 which angle gives the maximum speed for the angle $(\theta_0 = 90 ^{\circ})$. $L_1 = 0.4 m $ } \label{FCbTau}
\end{figure}

  In actual tennis stroke, most of the players use time lagged racket
movement. They intentionally keep the racket back as forearm rotates
then they start to move the racket to get a high angular velocity
$\omega_2$. The main difference of double pendulum when compared to
the single pendulum, is the separate movement of the first rod (arm
and forearm system) and the second rod (the racket system). In Fig.
\ref{FCbTau}, we plotted the speed of the rebound ball as a function
of increase in the torque $C_2$ and the delay time $\tau$. We set
the projected length in $xy$ plane  between the elbow and the hand
$L_1 $ as $0.4m$.  The torque applied to the racket at the joint of
two rods, varies from $0$ to $10N$. The time delay $\tau$ is the
starting time at which we applied the torque to the racket. After
waiting for $\tau$ second, the torque suddenly changes from zero to
a certain value $C_2$ till $0.5ms$ in our numerical calculation.
This time is far behind the contact time (around $0.2 ms$).
   At first glance, if we start to apply the torque early, the angular velocity of the
 racket may accelerate little bit more. It's simply because the
 acceleration of the angular velocity depends on the torque.
 If the torque  $C_2$ is less than $1.56N$, the maximum speed of
 the rebound ball is  obtained when the delay time $\tau$ is zero as we expected.
 However, for the higher torque $C_2>1.56N$,the numerical results in Fig. \ref{FCbTau} are different
 from our simple intuition. For a given torque value $C_2 > 1.56 N $, there
 always exist a certain time delay $\tau$  at which the speed of the
 rebound ball is maximum and the $\tau$ is not zero. In other
 words, the important thing to get high speed of the rebound is
 not the total amount of impulse (torque $\times$ (applied time)), but the timing
 when the torque starts.

\begin{figure}[htbp]
\centering
\includegraphics[width=5cm]{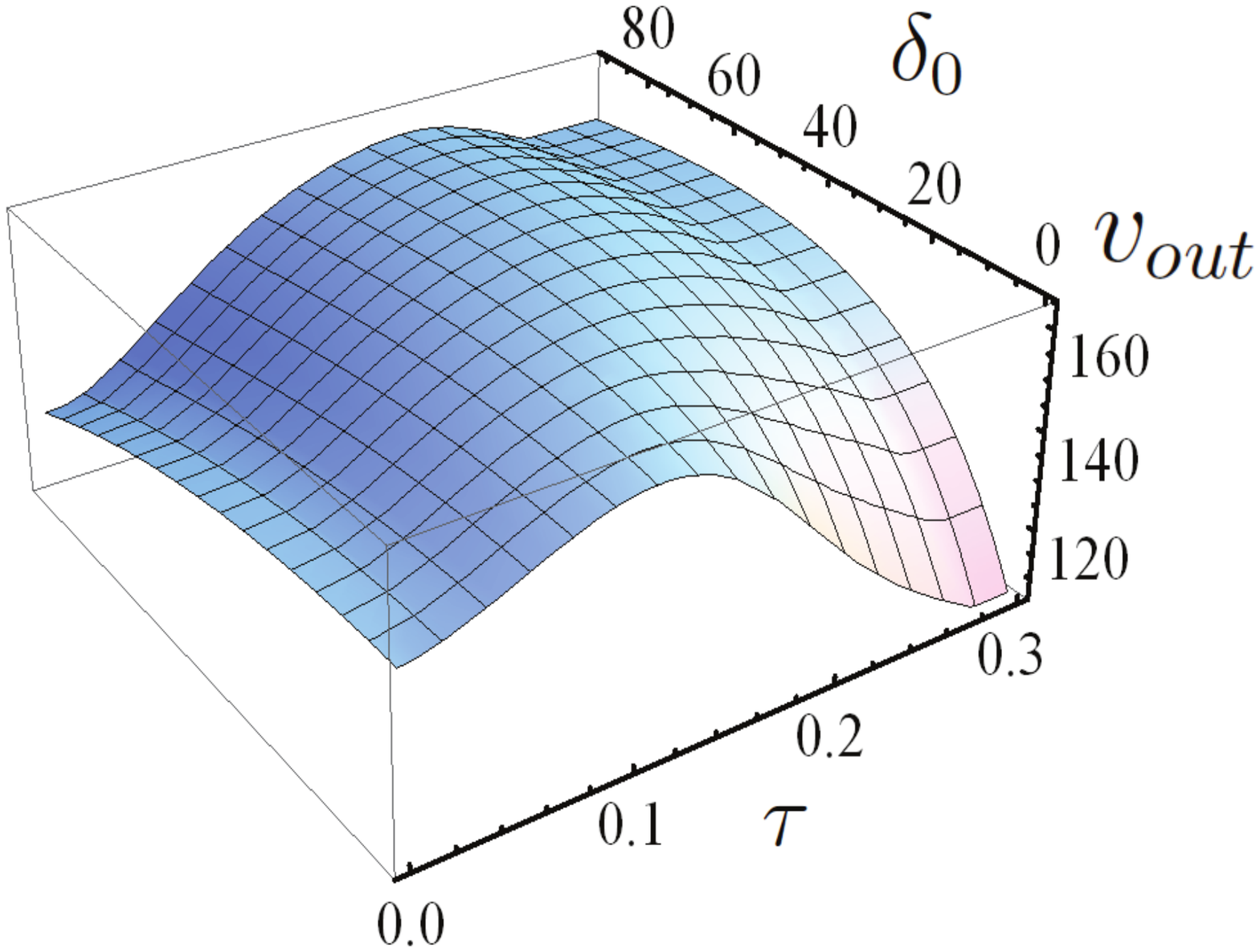}
\caption{ The speed of the rebound ball (km/h) as a function of the
time lag $\tau$ and initial angle $\delta_0$ with an initial angle
$\theta_0 = 90 ^{\circ}$} \label{FtimeDelta}
\end{figure}

\begin{figure}[htbp]
\centering
\includegraphics[width=5cm]{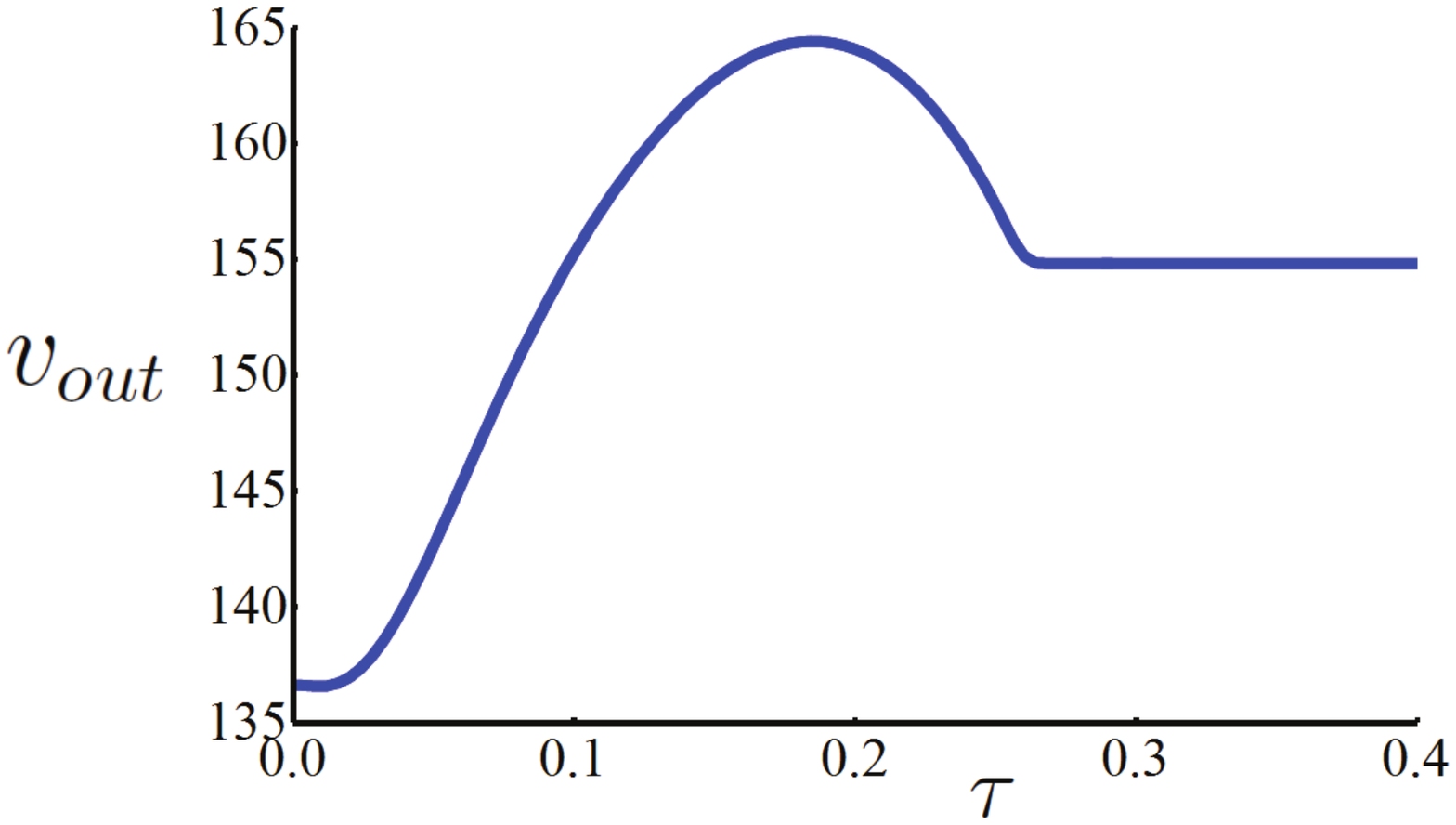}
\caption{The speed of the rebound ball as a simple function of time
delay $\tau$. At $\tau =0.185 s$, the $v_{out}$ has it's maximum .}
\label{FtimeDelay}
\end{figure}

\begin{figure}[htbp]
\centering
\includegraphics[width=5cm]{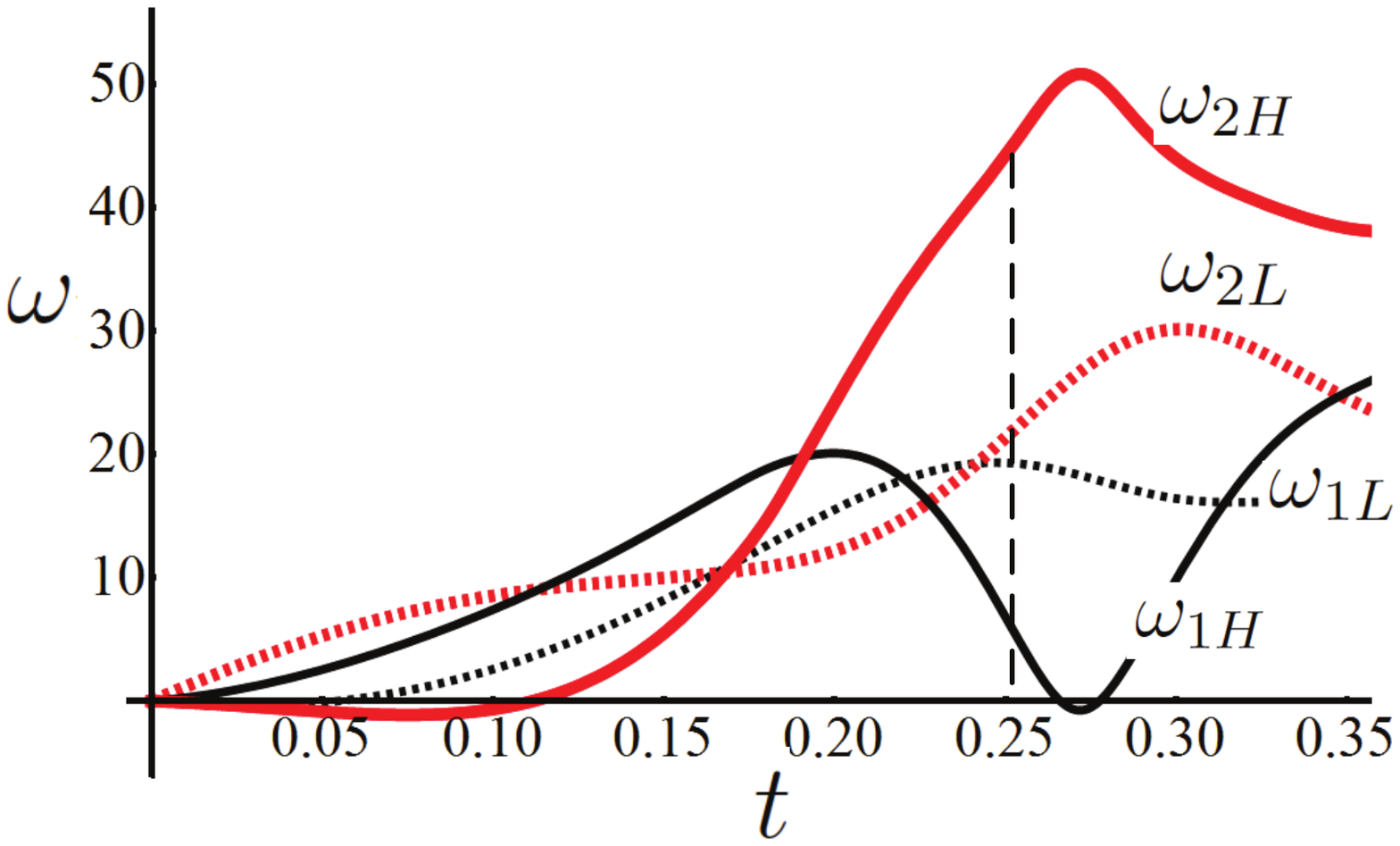}
\caption{ The angular velocity of the forearm system $\omega_1 $
  and  the racket system ($omega_2$). $(\omega_{1L}, \omega_{2L})$ indicates the angular velocities when
  the time delay is $\tau_L$, and  $(\omega_{1H}, \omega_{2H})$ indicates the angular velocities when
  the time delay is $\tau_H$} \label{FtimedelayW1W2}
\end{figure}

If the torque $C_2$ is lagged, the initial angle $\delta =26.1
^{\circ}$ to get a maximum speed of the rebound ball for the fixed
initial angle $\theta_0 = 90 ^{\circ}$ is no more the optimum
condition. In order to see the time lag effect clearly, we set the
torque  $C_2 =5 N$.
 In Fig. \ref{FtimeDelta}, we plotted the speed
of the rebound ball as a function of the time lag $\tau$ and initial
angle $\delta_0$ with an initial angle $\theta_0 = 90 ^{\circ}$. The
new optimum condition to get maximum speed of the rebound ball is
the time lag $\tau = 0.185 s$ and the initial angle $\delta_0 = 52.5
^{\circ}$.
In Fig. \ref{FtimeDelay}, we plotted the speed of the rebound ball
as a simple function of time delay $\tau$. If the time lag $\tau$ of
the torque $C_2$ is $\tau_L = 0$, the rebound ball speed is $v_L =
136.6 km/h$. And
 $v_{out}$  has its maximum ($v_H = 164.4 km/h$)
  at $\tau_H  = 0.185 s$. The speed increases by
  about $20 \% $ by adjusting the time delay with the same magnitude
  of the torque $C_2$. The angular velocity of the forearm system
  and  the racket system is shown in Fig. \ref{FtimedelayW1W2}. For two
  time delay $\tau_L$ and $\tau_H$, the difference in the contact time is less then $10ms$.
  The dashed line in  Fig. \ref{FtimedelayW1W2} indicates the time
  at which the ball collides with the racket.  We note that the racket hit
  the ball before the racket has its maximum angular velocity.  In
  earlier work \cite{RodCross11}, the authors have analyzed the double pendulum
  system in order to get the maximum angular velocity $\omega_2$.
  However, this may mislead the double pendulum system. Considering
   Fig. \ref{Fw1Nw2}, it is clear that higher angular velocity gives lower
    speed of the rebound ball.

  In Fig. \ref{FtimedelayW1W2}, when the angular velocity $\omega_{2H}$ has its maximum,
   the angular velocity $\omega_1$ is almost zero. This is explained
   for the double pendulum model of tennis stroke. The angular
   momentum and energy of the first rod (arm and forearm system) were totally  transferred
   to the racket system in order to get maximum angular velocity of the racket.

\begin{figure}[htbp]
\centering
\includegraphics[width=5cm]{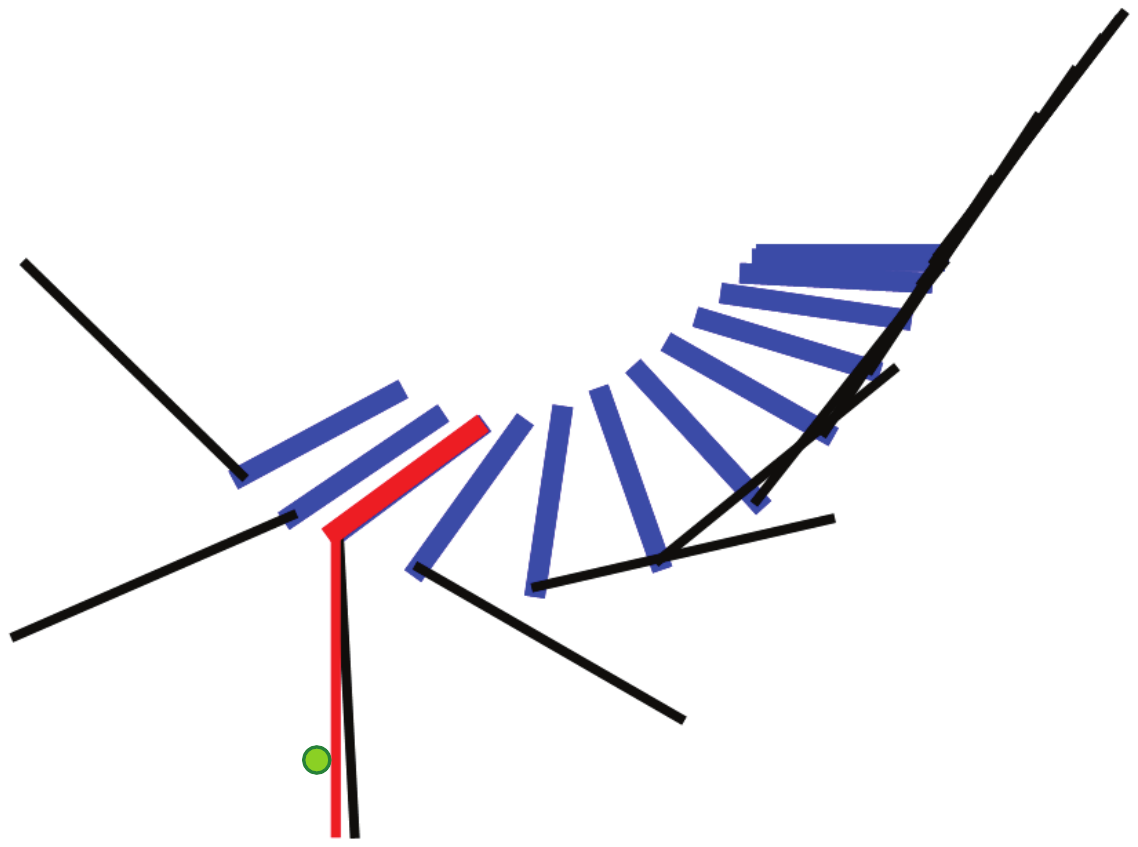}
\caption{Time dependent trace of the tennis racket system (black
bar) and the arm and forearm system (blue bar), when the time delay
is $\tau_H$. The motions are captured from $t=0$ to $t=0.32s$
evenly. The red one is the racket position at the contact time. }
\label{TimeLaggedStroke}
\end{figure}

\begin{figure}[htbp]
\centering
\includegraphics[width=5cm]{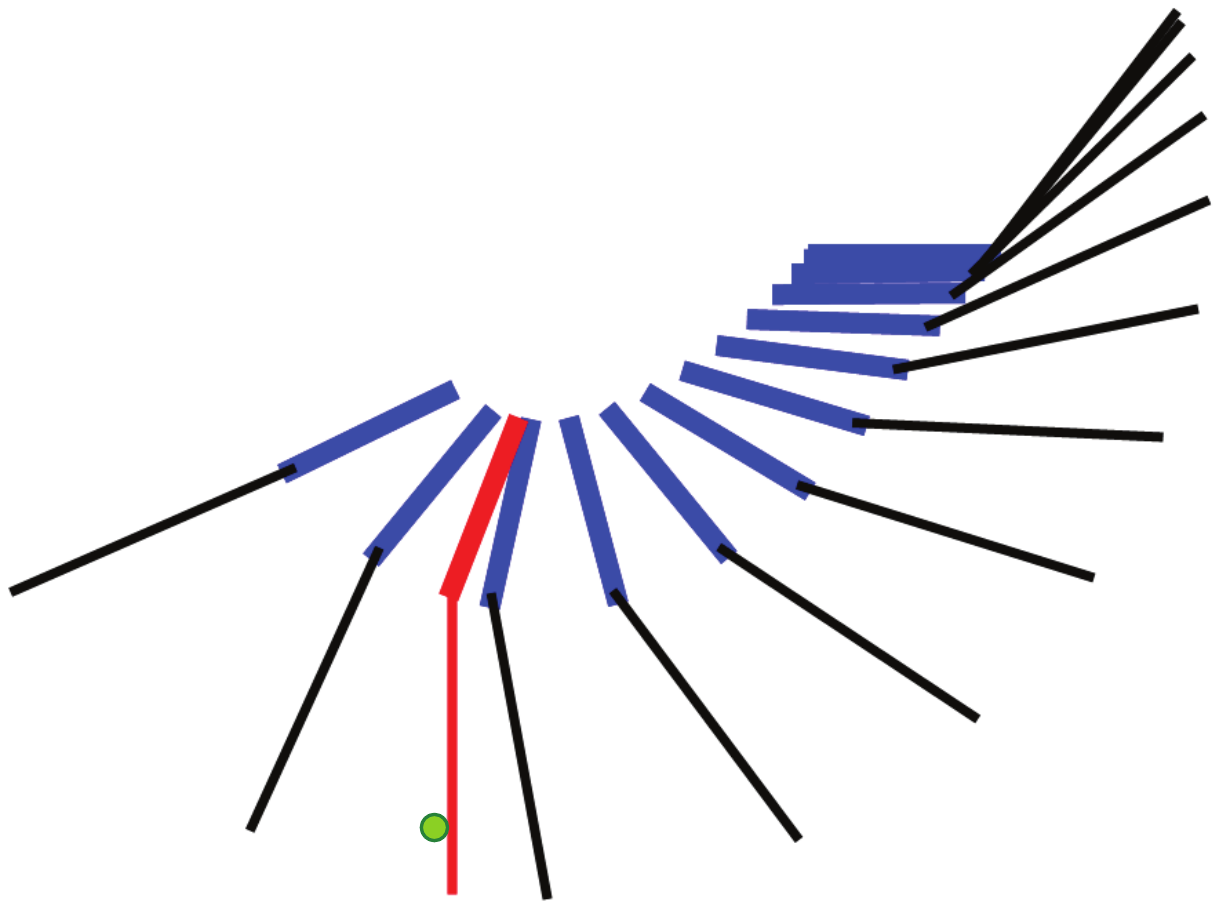}
\caption{Time dependent trace of the tennis racket system (black
bar) and the arm and forearm system (blue bar), when the time delay
is $\tau_L =0 $. The motions are captured from $t=0$ to $t=0.32s$
evenly. The red one is the racket position at the contact time.}
\label{ConstantStroke}
\end{figure}

We plotted the first rod (forearm and arm system) and the racket
system for the time delays represented as $\tau_H$ and $\tau_L$ in
Fig. \ref{TimeLaggedStroke} and \ref{ConstantStroke}, respectively.
When the first rod starts to rotate, the racket stayed back in Fig.
\ref{TimeLaggedStroke} till time $\tau_H$. After applying the torque
$C_2$, the racket suddenly starts to rotate, and hits the ball with
the angle $\theta_c  = 53.3 ^{\circ}$. The racket and the first rod
are shown in red at the contact time. Comparing the stroke with a
constant torque (Fig. \ref{ConstantStroke}), the racket rotates more
rapidly at the contact time.
  In double pendulum model for the stroke, this phenomena was
  expected qualitatively. In our model we quantitatively demonstrated that
  why time lagged stroke is needed and the extent to which the velocity can be increased.    ,


\section{Need more in order to increase the speed of the rebound ball.}

\begin{figure}[htbp]
\centering
\includegraphics[width=5cm]{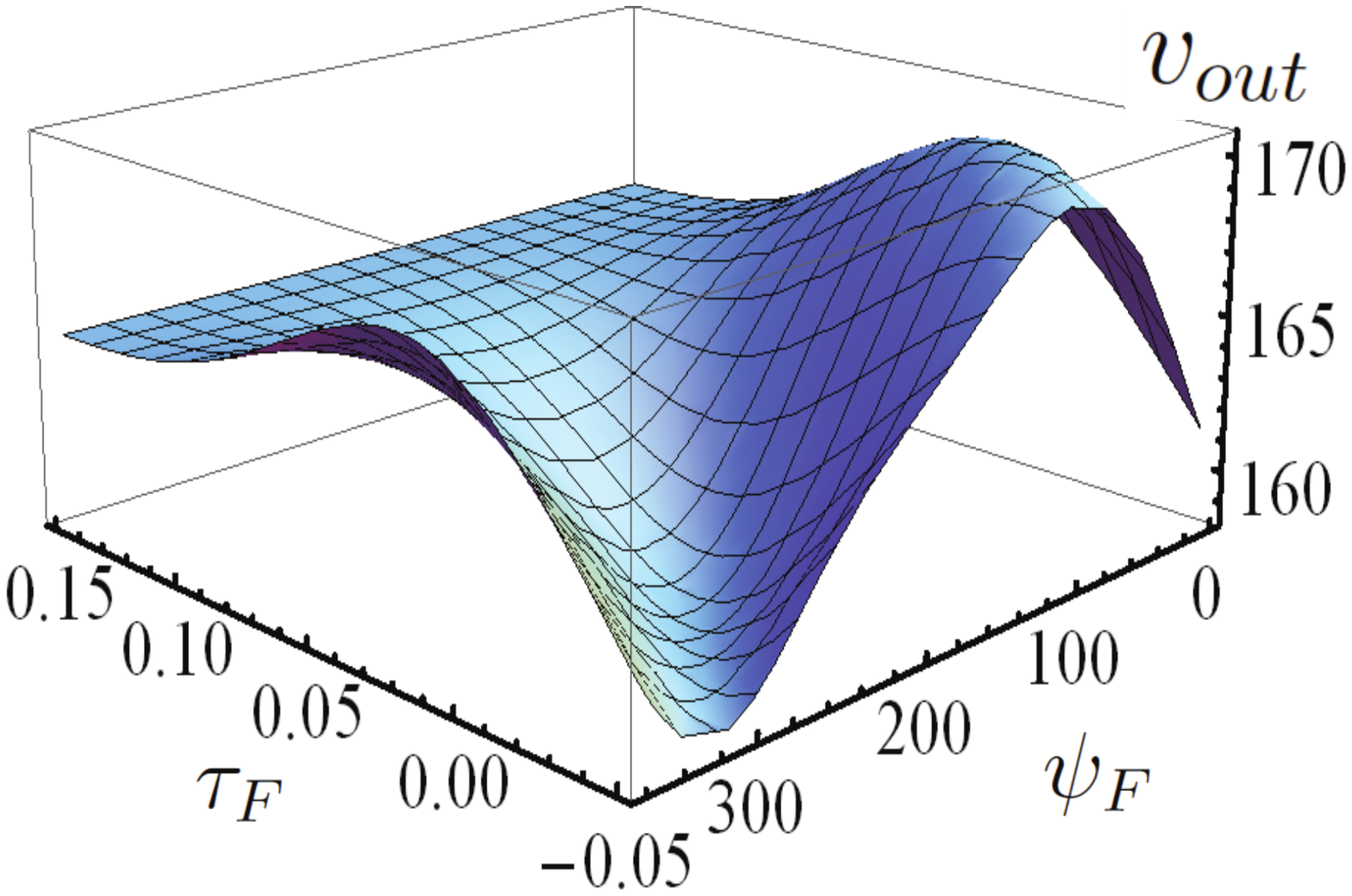}
\caption{The speed of the rebound ball as  a functions of the extra
force with a direction of $\Psi_F$ and with a extra time delay
$\tau_F$ } \label{FForceM}
\end{figure}

\begin{figure}
\includegraphics[width=5cm]{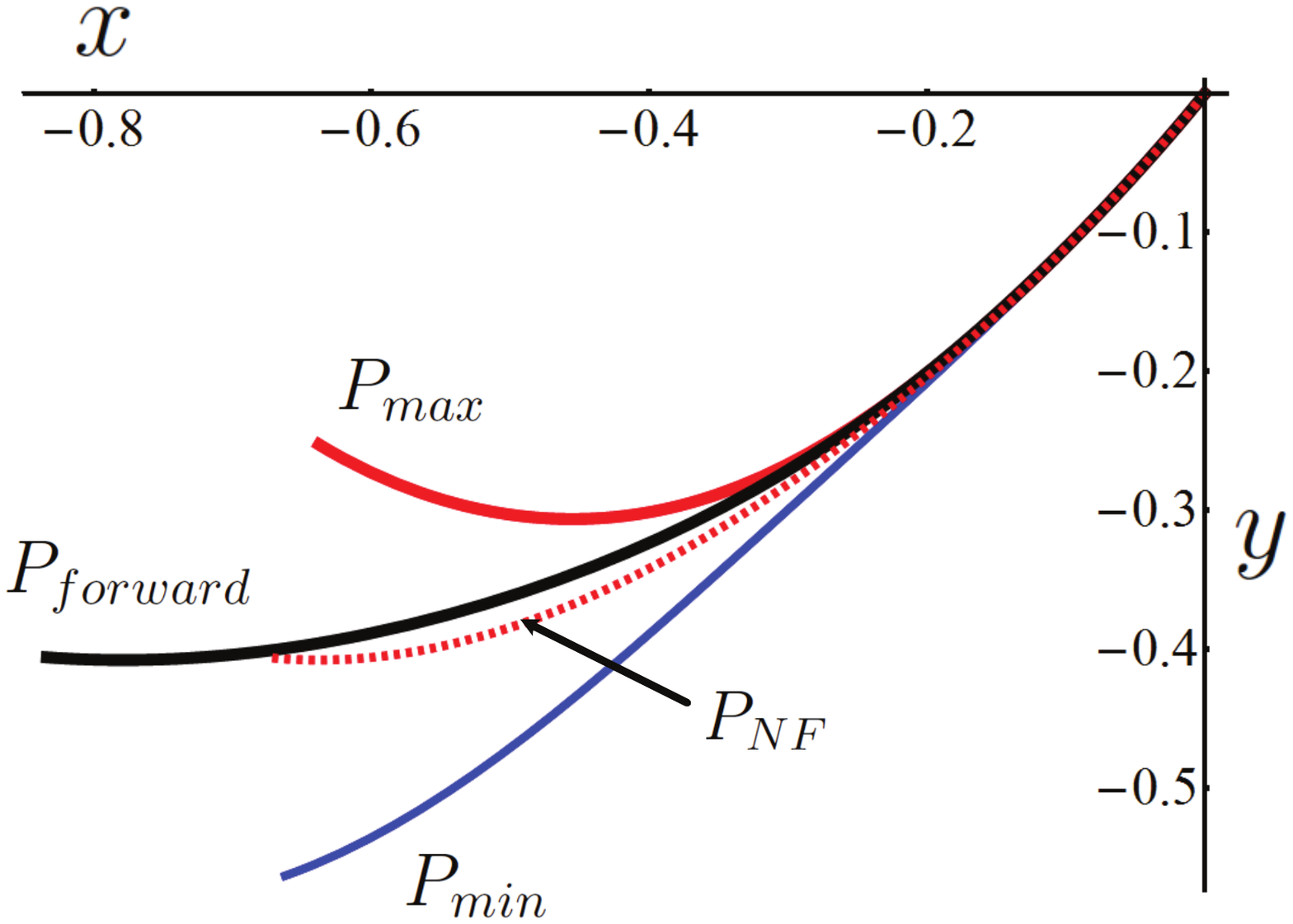}
\caption{The trajectory of the elbow for four cases from the time
$t=0$ till the contact time. $P_{NF}$ is the path when no additional
force is added. $P_{forward}, P_{max}, P_{min}$ are elbow's
trajectory when the direction of the force $\Psi_F$'s are $
180^{\circ}, 76^{\circ}, 275^{\circ}$, respectively.} \label{FArmM}
\end{figure}

In the previous section, we noted that the important thing is not
the total impulse to accelerate the racket, but the timing in order
to increase the speed of the rebound ball. Now, we analyze the
influence  of the  elbow movement on the speed of the rebound ball.
We made a simple model to add an additional movement of the elbow.
The added force around the time $\tau_H$  as follows,
\begin{eqnarray}
  F_x (t) = F_0 \cos \Psi_F  \exp (-(t- (\tau_H +\tau_F))^2/ t_d^2 )   \nonumber \\
  F_y (t) = F_0 \sin \Psi_F  \exp (-(t- (\tau_H +\tau_F))^2/ t_d^2 )  \label {Force}
\end{eqnarray}
where $\Psi_F$ is the angle measured from the $+x$ axis
counterclockwise, we also set $F_0 = 20 N $, and the width $t_d =
0.05s$. Since the ball is rebound to $-x$ axis, the proper direction
of the force seems to be in the $-x$ direction. We plotted the speed
of the rebound ball when the extra force is applied in the direction
of $\Psi_F$ and with extra time delay $\tau_F$ in Fig.
\ref{FForceM}. Numerical results show that we can obtain the maximum
speed of the rebound ball when the angle is $76 ^{\circ}$. When the
direction of the force $\Psi_F = 180 ^{\circ}$, the speed of the
rebound ball is almost same to the speed of the ball without the
additional force. In other words, the external movement towards $-x$
direction does not give any additional speed to the rebound ball. In
Fig. \ref{FArmM}, we plotted the trajectory of the elbow from the
time $t=0$ till the contact time. $P_{NF}$ is the path when no
additional force is added. $ P_{max}, P_{min}$ are elbow's
trajectory when the direction of the force $\Psi_F$ are $76
 ^{\circ}$ and, $ 275 ^{\circ}$, respectively. Table \ref{Tb1}
  demonstrates the $v_{out}$
and the angular velocity $\omega_2$ for 4-cases. The movement with
$\Psi_F = 76 ^{\circ}$ is slightly backward and mainly perpendicular
to the direction of the rebound ball. From this result, we conclude
that the main factor for the increased speed of the rebound ball is
not the linear momentum added to the racket but the angular momentum
added to the racket system.

\begin{table}
\centering \caption{Additional Force dependent output} \label{Tb1}
\begin{tabular}{c|c|c|c|r}
\hline
 $F_0$ (N)&$\Psi_F (^{\circ}) $  & $ v_{out}$ (km/h)  & $\omega_2 $ $rad/s$  \\ \hline \hline
20&180 & 163.3 &39.1  \\ \hline 20& 275 & 158.8 &39.2  \\ \hline
 20& 76 & 170.6 &52.2  \\ \hline 0& 0 & 164.4 &44.7  \\ \hline
     \hline
\end{tabular}
\end{table}

\section{Conclusion and Discussion.}

   The double pendulum model is applied to the baseball, tennis, and
golf. It analyzes the swing pattern to maximize the angular velocity
of the hitting rod such as racket, bat, and club on the assumption
that the angular velocity is the dominant factor to attain speed of
the rebound ball. If we set, $\theta_c$,  the angle of the first rod
at the impact time, it seems to be obviously reasonable.  On the
other hands, if we release the angle $\theta_c$, the speed of the
rebound ball is not a simple function of the angular speed of the
hitting rod.  The speed of the rebound ball is different even though
the angular velocities at the contact time are same in Fig.
\ref{Fw1Nw2},. Furthermore, considering the whole stroke, the
maximum angular velocity $\omega_2$  for the lower speed of the
rebound ball is greater than the maximum angular velocity $\omega_2$
for the higher speed of the rebound ball. Therefore, to attain
maximum speed of the rebound ball, it's not sufficient to set the
condition to generate high angular velocity $\omega_2$ of the
hitting rod.

  In the double pendulum, the efficient way to generate high angular velocity of the hitting rod
  is the energy transfer from the first rod to the second hitting
  rod. In other words, when the angular velocity $\omega_1$ of the forearm
  system is zero,
  the angular velocity of the hitting racket has maximum value as shown in
  Fig. \ref{FtimedelayW1W2}. We analyzed the time lagged torque
  effect for the double pendulum system. With applying constant torques  $C_1 , C_2 $ on the forearm system and on the racket,
  respectively, the speed of the rebound ball can be calculated.
  For the same condition, if we simply  hold the racket for a short time without enforcing a torque
  $C_2$, then applying the torque $C_2$ at the proper time ($t =
  \tau_H $), the speed of the rebound ball increases by $20 \%$ as
  can be seen in Fig. \ref{FtimeDelay}. The reason is mainly because the double pendulum
  system is not a simple linear system. Adding the energy to the
  double pendulum system does not directly increase the speed of the
  rebound ball.

We also analyzed the elbow movement effect. In addition to the
velocity of the elbow for the medium pace forehand, we added extra
movement of the elbow. At a first glance, if we add extra movement
towards the rebound ball's direction, the speed of the ball is
increased. But the double pendulum system is not a simple linear
system. When the direction of the elbow movement is perpendicular to
the ball's direction, the speed of the rebound increases. Actually
the direction is towards the center of the elbow's circular
movement. In other words, the added centripetal force does not add
not the linear momentum of the racket, but the angular velocity of
the racket.

  Although our collision model is applied in one dimension,
 this collision process allows us to analyze the double pendulum system
 in a more realistic manner. We showed that the speed of the rebound ball does not
 simply depend on the angular velocity of the racket. The increase in the ball speed
 by the proper time lagged racket rotation was numerically studied.
 The elbow movement for adding the ball's speed was counter
 intuitive.  The addition of simple linear momentum to the elbow is
 not important; however, the elbow should move in order to add angular
velocity to the racket.

   In actual tennis stroke, the motion occurs in three dimensions and
   the  magnitudes of the forces and torques are dependent on the
muscle shape and movement. We did not include any bio-mechanical
information such as pronation and we did not include the spin of the
ball in any way. The numerical data may also be not suitable for
some players. However, our study on the double pendulum system for
tennis stroke provides some insights to attain an efficient way to
stroke a tennis ball.

%
%
%

%

\acknowledgements

 This study was supported by the Basic Science
Research Program through the National Research Foundation of Korea
(NRF) funded by the Ministry of Education, Science and Technology
(NRF-2014R1A1A2055454)

%
%


\begin{references}


\bibitem{chaos1} R.B. Levien, and S. M. Tan, "Double pendulum: An
experiment in chaos" Am. J. Phys. {\bf 61}, 1038-1044 (1993) 1T.
\bibitem{chaos2} T. Shinbrot, C. Grebogi, J. Wisdom, and J. A. Yorke, "Chaos in a
double pendulum", Am. J. Phys. {\bf 60}, 491-499 (1992)
\bibitem{chaos3} G. Vadai, Z. Gingl and J. Mellar, "Real-time demonstration of the main
characteristics of chaos in the motion of a real double pendulum",
Eur. J. Phy. {\bf 33}, 907 (2012)

\bibitem{Ref3} D. Williams, "The dynamics of the golf swing.", Q.
J. Mech. Appl. Math. {\bf 20}, 247-264 (1967)
\bibitem{Ref4} C. B. Daish, {\it The Physics of Ball Games "}.
(English University Press, London, 1972)
\bibitem{Ref5} T. Jorgensen, "On the dynamics of the swing of a golf club,"
Am. J. Phys. {\bf 38}, 644-651 (1970)
\bibitem{Ref6} T. Jorgensen, {\it The Physics of Golf} (Springer-Verlag, New York, 1999), 2nd ed.


\bibitem{RodCross05} R. Cross, "A double pendulum swing experiment: In search of a better bat,"
 Am. J. Phys. {\bf 73} , 330-339 (2005).
\bibitem{RodCross11} R. Cross, "A double pendulum model of tennis
strokes", Am. J. Phys. {\bf 79} (5) , 470 (2011).
\bibitem{RodCross00} R. Cross,"The coefficient of restitution for collisions of happy balls,
unhappy balls, and tennis balls",  Am. J. Phys. {\bf 68} , 1025-1031
(2000).

\bibitem{RodCrossImpact}  R. Cross, "Impact of a ball with a bat or racket, " Am. J.
Phys. 67, 692-702 (1999)
\bibitem{ObliqueImpact}  R. Cross, "Oblique impact of a tennis ball
on the strings of a tennis racket", Sports Engineering,  {\bf 6},
235-254 (2003)


\bibitem{impact} R. Cross, "Impact of a ball with a bat or racket," Am. J. Phys. {\bf 67}, 692-702 (1999)

\bibitem{wristTorque} E. J. Sprigings and R. J. Neal, "An insight into the importance of wrist
torque in driving the golfball: A simulation study," J. Appl.
Biomech. {\bf 16}, 356-366 (2000)


\bibitem{swingweight} R. Cross and R. Bower, "Effects of swing-weight on swing speed
and racket power," J. Sports Sci. 24, 23?30 (2006).

\

\end{references}
\end{document}